# Effects of proton irradiation and ageing on the superconducting properties of single crystalline and polycrystalline MgB2.

G. K. Perkins[1], Y. Bugoslavsky[1], A . D. Caplin[1], J.Moore[1], T. J. Tate[1], and R. Gwilliam[2], J.Jun[3], S.M.Kazakov[3], J. Karpinski[3] and L. F. Cohen[1]

[1] Centre for Electronic Materials and Devices, Blackett Laboratory, Imperial College, London SW7 2BZ, UK

[2] EPSRC Ion Beam Centre, University of Surrey, Guildford, Surrey GU2 7RX, UK

3 Solid State Physics Laboratory, ETH, CH-8093 Zürich, Switzerland.

*Abstract*

We present magnetisation data on crystalline and polycrystalline $MgB_2$ before and after irradiation with ~1MeV protons. In the virgin crystal the critical current density is below our noise floor of $10^3 A/cm^2$ over most of the field range. However, after irradiation a peak occurs in the current density as a function of applied magnetic field as the upper critical field $H_{c2}$ is approached. After subsequent ageing over a time period of three months, the peak effect is greatly enhanced, exhibiting much stronger pinning over a wide field range, and the upper critical field is approximately doubled, accompanied by a 2K reduction in transition temperature. Similar studies were made on polycrystalline fragments, where irradiation leads to an increased irreversibility field ($J_c$ is enhanced at high fields but decreased at low fields) and a suppression in transition temperature. However, after two years of ageing both parameters returned towards those of the virgin sample.

## 1.1 Introduction

Artificial damage induced by irradiation provides a unique opportunity to study the effects of disorder on the electronic properties of superconducting materials. In general, increasing the levels of disorder can improve the critical current density $J_C$ and the irreversibility field $H_{irr}(T)$ by introducing more effective pinning centres into the system. Additionally, the disorder may affect electron scattering mechanisms, thereby altering the thermodynamic superconducting properties, in particular reducing the superconducting coherence length $\xi_{eff}$, and consequently increasing the upper critical field $H_{c2}$.

Most irradiation studies on $MgB_2$ so far have been carried out on polycrystalline samples, where enhancements in $J_c$ and $H_{irr}$ have been observed after heavy ion irradiation[1,2] and to a lesser degree with proton irradiation.[3] It has also been shown that damage by fast neutron irradiation can enhance $H_{c2}$ at the expense of a reduced $T_c$.[4] At higher neutron fluences both $T_c$ ad $H_{c2}$ are reduced considerably but can be restored after annealing at 700°C,[5] indicating that atomic disorder is relatively mobile in $MgB_2$.

Recently we observed a substantial enhancement $H_{irr}(T)$ in polycrystalline (PX) $MgB_2$ fragments after low energy proton irradiation.[6] However the mechanism by which this occurs, and the exact nature of the disorder is still unclear. In this paper we study the effect of similar proton irradiation on the properties of high quality $MgB_2$ single crystals and compare the results to those obtained earlier on the PX fragments. Additionally we study the effect of sample ageing on both systems, and hence gauge the mobility of the disorder created by the irradiation.

## 1.2 *Experimental*

Magnetisation loops of a high quality MgB$_2$ single crystal, having a transition temperature $T_c$~38K (onset) and a transition width ~1K, were measured using a Vibrating Hall Micromagnetometer (VHM). This device has exceptionally low background signals and high sensitivity for small samples, so that the upper critical field $H_{c2}$ can be reliably determined in this crystal (of order 200μm in size) by a well defined onset of reversible magnetisation at the transition (a detailed characterisation of similar crystals using this technique is described in ref.[7]). All of the crystal measurements reported here were performed with the applied magnetic field $H$ aligned parallel to the sample c-axis.

For the PX sample, several fragments of size ~300μm were selected from commercial MgB$_2$ powder (Alfa Aesar Company, 98% purity). They had slightly lower $T_c$~37K (onset) and a less sharp transition width~2K. As each fragment comprises randomly-oriented grains (around 5μm in size) $H_{c2}$ was not measured, and magnetisation loops were obtained using a conventional Vibrating Sample Magnetometer (Oxford instruments 3001).

Proton irradiation was carried out at the EPSRC Ion Beam Centre in the UK. In order to create fairly uniform damage through the depth of the samples, 15 consecutive implants were made with the beam energy varied between 400 keV and 2 MeV so that average damage is 1% displacements per atom in both samples. Further details of the irradiation procedure are given in ref. [6].

After the initial measurements the samples were stored at room temperature in a water-free environment. In order to study the effects of ageing, the crystal was remeasured after a period of 3 months, and the PX fragments after a period of 24 months.

All data sets are interpreted in the terms of the critical current density $J_C(B)$ calculated from the magnetic hysteresis using the standard critical state model.

## 1.3 Results

In the following section we present data for the both the crystal and PX samples in the virgin state and shortly after irradiation (within two weeks). We then proceed to discuss the effects of ageing.

### 1.3.1 Effects of Initial irradiation

Figure 1 shows $J_C(B)$ at 20K for both the crystal and the PX fragments before and after irradiation. For the virgin crystal $J_C(B)$ rapidly decreases with increasing field and we define the irreversibility field $H_{irr}(T)$ as the point at which $J_C(B)$ first falls below the noise floor of $10^3$ A/cm$^2$. The temperature dependences of $H_{irr}(T)$ for the two samples are shown in fig.2, and for the virgin crystal is much lower than the PX sample, reflecting the high degree of purity of the crystal.

After irradiation, the low field behaviour in the crystal is little changed. This is particularly true for temperatures above 20K where it can be seen in Fig.2 that the two $H_{irr}(T)$ dependences (before and after irradiation) track each other closely. At lower temperatures there is a slight enhancement in $H_{irr}(T)$ after irradiation. However, the most significant effects of irradiation are observed at higher fields (approximately $0.7H_{c2}$ and above) where after irradiation $J_C(B)$ exhibits a pronounced, and narrow, peak effect, rising sharply to a maximum of ~$3 \times 10^4 A/cm^2$ before decreasing to zero as $H_{c2}$ is approached. Similar peak effects have been observed in as grown $MgB_2$ crystals,[8] but to our knowledge this is the first time the peak effect has been induced artificially by irradiation. Note that neither $H_{c2}$ nor $T_c$ of the crystal are visibly-affected by the initial irradiation (fig.3).

In comparison the effect of irradiation on the PX sample is to increase $H_{irr}(T)$ (enhancing $J_C(B)$ at high fields) although at lower fields $J_C(B)$ is depressed. Additionally $T_c$ is reduced, with an onset of 33K and a transition width of 10K.[6]

### 1.3.2 Effects of Ageing

The $J_C(B)$ data at 20K taken after ageing are shown in fig.4 and clearly demonstrate that ageing of the crystal sample for three months greatly enhances the peak effect. It is approximately twice as large in magnitude and very much broader. More significantly, the upper critical field has a much steeper temperature dependence (fig.3) and is about double the un-aged value at 20K. However, there is now a marked depression in $T_c$ of about 2K ($T_c$~36K). Note that the irreversibility line is now similar to that observed in the virgin fragments (fig.2). In contrast, the effect of 24 months ageing in the PX is to reduce $J_C(B)$ and $H_{irr}(T)$ (figs.2,4) so that the system more closely resembles that of the virgin PX. This is also supported by a recovery of $T_c$ to ~36K.

### 1.4 Discussion

Although irradiation of the crystal clearly introduces disorder into the system, it is not sufficiently strong to cause significant changes in $J_C$ at low magnetic fields. At low temperatures irradiation produces a slight enhancement in $H_{irr}(T)$, but above 20K there is no observable effect. This scenario makes sense if the predominant pinning mechanism in both case (before and after irradiation) for temperatures above 20K arises from surface or geometric barriers (as suggested by the form of the magnetisation loops[7]), and bulk pinning only comes into play at lower temperatures (hence the increased bulk pinning due to irradiation is only evident at low temperatures).

On the other hand, at high magnetic fields the irradiation of the crystal has a pronounced effect across the whole temperature range. In particular the appearance of the peak can be understood within the framework of bulk collective pinning and of the properties of the vortex lattice VL. The earliest analysis[9] argues that as $H_{c2}$ is approached there is a softening of the vortex lattice shear modulus $c_{66}$ allowing the VL to accommodate better to the underlying disorder. More recent work[10,11,12] suggests that the softening of the VL arises from an ordered to disordered transition within the VL, possibly due to plasticity associated with the nucleation and subsequent motion of vortex lattice defects. The fact that irradiation of our crystal does not produce an observable current density at lower fields suggests that elastic interactions in the VL dominate over the pinning interactions, leading to long vortex correlation lengths and hence weak pinning. However, because the energy for lattice defect nucleation should decrease with increasing field,[13,14,15] eventually (i.e. around the onset of the peak effect) the disorder is strong enough to nucleate defects, ultimately giving rise to the increase in $J_C$. This clearly demonstrates that the underlying pinning centres are very much more effective once the VL disorders.

Although the microscopic mechanism is as yet unclear, the ageing process clearly increases the pinning strength considerably, and despite the enhancement in $H_{c2}$, the onset of the peak effect moves to *lower* magnetic fields, consistent with stronger pinning forces now being better able to nucleate disorder in the VL. It has recently been shown that the extent of the ordered VL phase (in the *H-T* vortex phase diagram) can be highly sensitive to the underlying disorder.[16] However, as we see greatly enhanced pinning in both VL phases after ageing, the indications are that the pinning energies are much stronger in the aged crystal. Additionally, the ageing process appears to have reduced the electron mean free path as indicated by the observed enhancement of $H_{c2}$, albeit at the expense of a reduced $T_c$.

In contrast, for the PX sample the initial irradiation leads to a decrease in $J_C$ at low fields, and an enhanced $J_C$ at higher fields (in some ways this is similar to the observed behaviour in the crystals, i.e irradiation has the greatest beneficial effect at high fields). $H_{irr}(T)$ is always higher in the PX than in the crystal, but a direct comparison is misleading due to anisotropy and random alignment of the grains within the PX, which should tend to enhance both $H_{irr}(T)$ and $H_{c2}(T)$.

Although in the crystal sample the peak effect (particularly after ageing) yields high $J_c$ values which are comparable to the PX data (fig.4), the form of $J_c(B)$ is entirely different in the PX sample (i.e. there is no observable peak effect). This may be due to the random orientation of the grains within each fragment, or to the role of the grain boundaries (which appear to act as pinning centres in $MgB_2$ thin films[17]) on the pinning mechanism. From topological considerations alone, it is evident that grain boundary pinning constitutes an entirely different pinning regime to that of pinning within the grains.

Even though the pinning processes may be very different in the PX sample, it is perhaps surprising that the effect of ageing is generally opposite to that in the crystal. Rather than increasing the disorder, ageing in the PX appears to shift the properties towards those of the virgin state (born out by the reduction of $H_{irr}(T)$ and the increase of $T_c$). The key question is that of the nature of the irradiation induced disorder and its mobility in $MgB_2$. The most likely forms of disorder resulting from proton irradiation are vacancies (created by collisions between the protons and the crystal lattice) and the protons themselves, which are most likely incorporated into the boron layer, attached to the spare boron bond. The ageing effect that we have demonstrated here suggests that the disorder is relatively mobile within $MgB_2$ and probably reflects diffusion of the vacancies. The role the of grain boundaries in the PX must also be considered; diffusion within the grains takes place over relatively short length scales (and should therefore be fast), so that diffusion along the grain boundaries plays a major role, most likely leading to faster diffusion rates than in the crystal (diffusion along grain boundaries is commonly several orders of magnitude faster than within the bulk). This mechanism could allow the disorder to relax fully out of the system comparatively quickly in the PX, thereby returning the system to the virgin state (vacancies may also annihilate or become less effective at the grain boundaries). In the crystals the diffusion initially leads to a more disordered state (i.e. stronger pinned), possibly associated with vacancy clustering, making larger, more effective pinning centres. It should be pointed out that such a comparison may be distorted due to the different ageing periods in the two cases, however we have remeasured the crystalline sample after total ageing period of 6 months and observed no further significant change, suggesting that time alone is not responsible for the different effects of ageing in the crystal and PX.

As a function of disorder we observe an anti-correlation between $T_c$ and $H_{c2}$ (when $T_c$ is enhanced there is a corresponding drop in $H_{c2}$, and vice versa) which can be understood by the two-gap nature of the superconductivity in $MgB_2$. However, $T_c$ is expected to decrease only if the disorder increases the interband electronic scattering rate.[18] The fact that in the single crystal we do not see any suppression in $T_c$ or enhancement in $H_{c2}$ shortly after the irradiation suggests that the induced disorder does not initially enhance interband scattering, but does so after the ageing process.

## *1.5 Conclusions*

We have studied the comparative effects of proton irradiation in $MgB_2$ single crystals and PX fragments. Irradiation of the crystal leads to a peak effect in $J_C(B)$ as $H_{c2}$ is approached. After an ageing period of three months the crystal exhibits a significantly enhanced peak effect. Additionally, $H_{c2}$ is approximately doubled at low temperatures and the critical temperature is suppressed by 2K. In the PX fragments, the initial irradiation enhances the irreversibility line and suppresses $T_c$, but after ageing the properties revert towards those of the virgin state.

In terms of applications it is encouraging to note that the aged defects provide pinning of a strength relevant to high current conductors, i.e. $3 \times 10^4 A/cm^2$ at 20K, and 2T. Furthermore, given that the defects were formed as a consequence of irradiation damage at a level of 1% d.p.a. their volume fraction in our crystal must be quite small. Detailed microstructural studies will be needed to establish the nature of these useful defects.

**Acknowledgement**

We thank Prof. A. M. Stoneham for helpful discussions as to the likely nature of defects in proton irradiated $MgB_2$. This work was funded by the UK Engineering and Physical Sciences Research Council grant number GR/N29310/01

Figure 1. $J_c(B)$ for the crystal and PX samples in the virgin state and shortly after proton irradiation at 20K. For the irradiated crystal, at low fields $J_c(B)$ is similar to the virgin sample and $J_c$ first falls below the noise floor at ~0.2T (first closure). At higher field the peak effect forms, with the current falling below the noise floor for a second time at ~2T (second closure). For the PX sample, the irradiation suppresses $J_c$ at low fields but enhances it at high fields, consequently increasing the irreversibility field.

Figure 2. $H_{irr}(T)$ for the crystal and PX samples in the virgin state, shortly after irradiation and after a subsequent ageing. The irradiated crystal has two $H_{irr}$ values corresponding to first and second closure of the magnetic hysteresis (fig.1). $H_{irr}(T)$ (first closure) is similar to the virgin crystal, while $H_{irr}(T)$ (second closure) closely tracks $H_{c2}(T)$. After 3 months ageing there is only one closure of the hysteresis and $H_{irr}(T)$ is substantially higher than that of the second closure in the un-aged sample. For the PX sample irradiation enhances $H_{irr}(T)$ (approximately by a factor 2) while after 24 months ageing $H_{irr}(T)$ is suppressed, approaching that of the virgin sample.

Figure 3. $H_{c2}(T)$ for the crystal in the virgin state, shortly after irradiation and after 3 months ageing. The initial irradiation makes little difference to $H_{c2}$ and $T_c$. However after ageing, $T_c$ is slightly suppressed (by~2K) and $H_{c2}(T)$ becomes much steeper, and more than doubles in value at low temperatures. The inset shows the behaviour near to $T_c$ on an expanded scale.

Figure 4. $J_c(B)$ for the crystal and PX samples shortly after irradiation and after ageing at 20K. For the crystal ageing greatly enhances the peak effect, pushing $J_c(B)$ above that of the PX sample over a significant field range. In the PX sample ageing has the opposite effect, reducing both $J_c(B)$ and the irreversibility field.

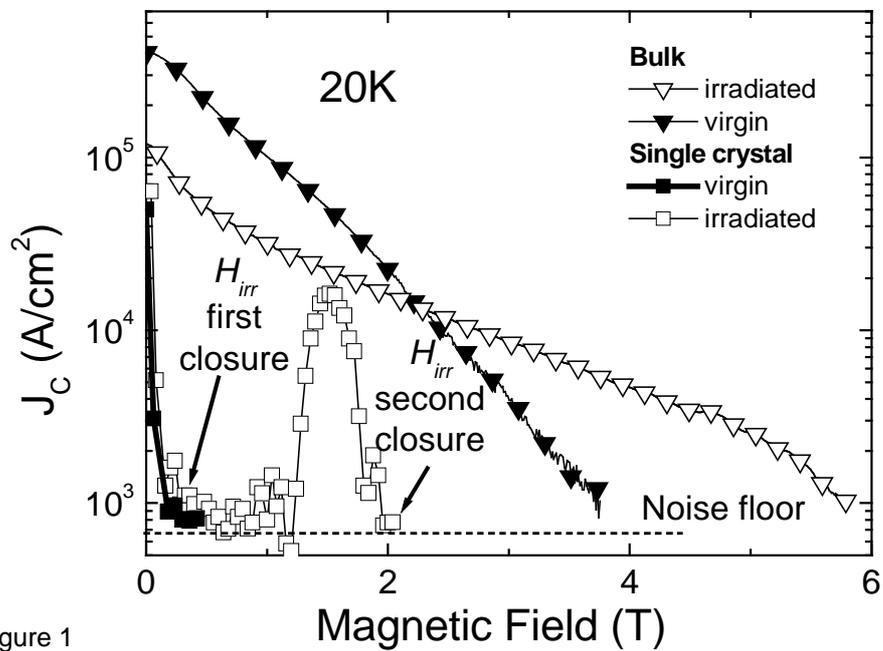

Figure 1

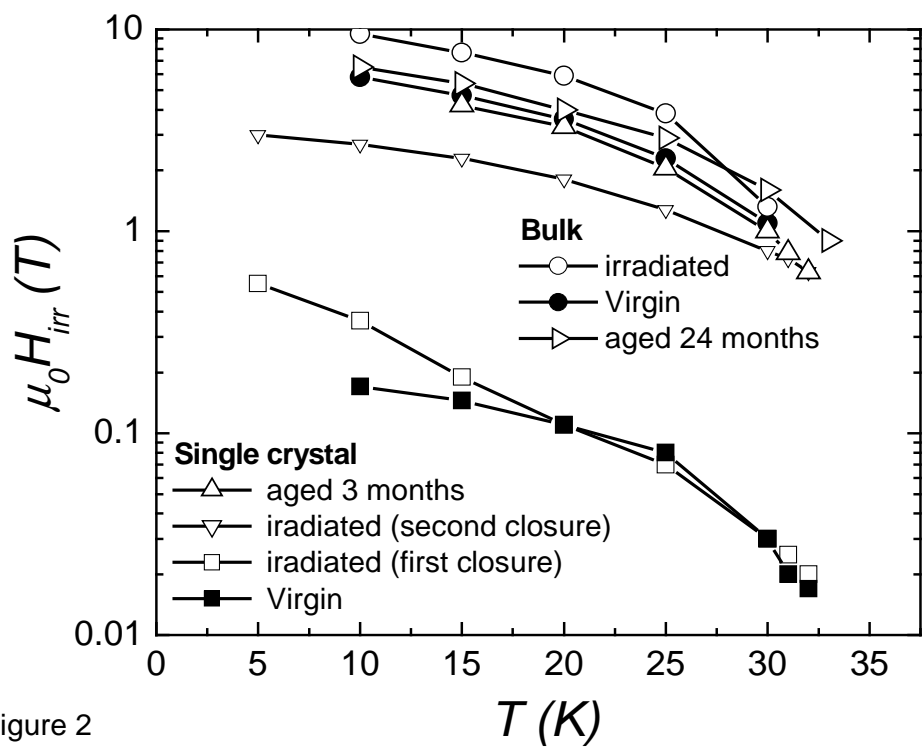

Figure 2

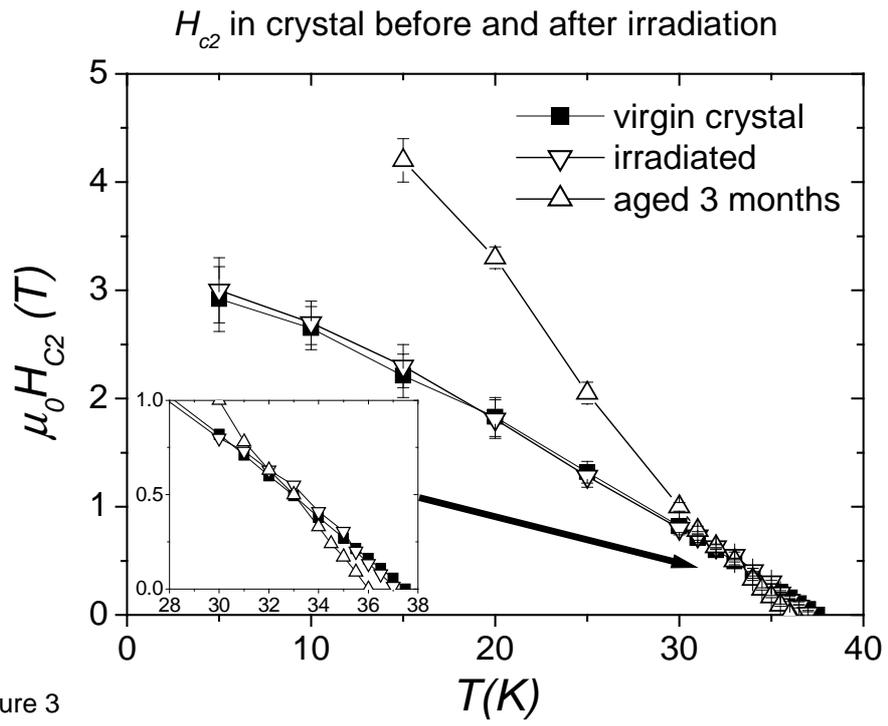

Figure 3

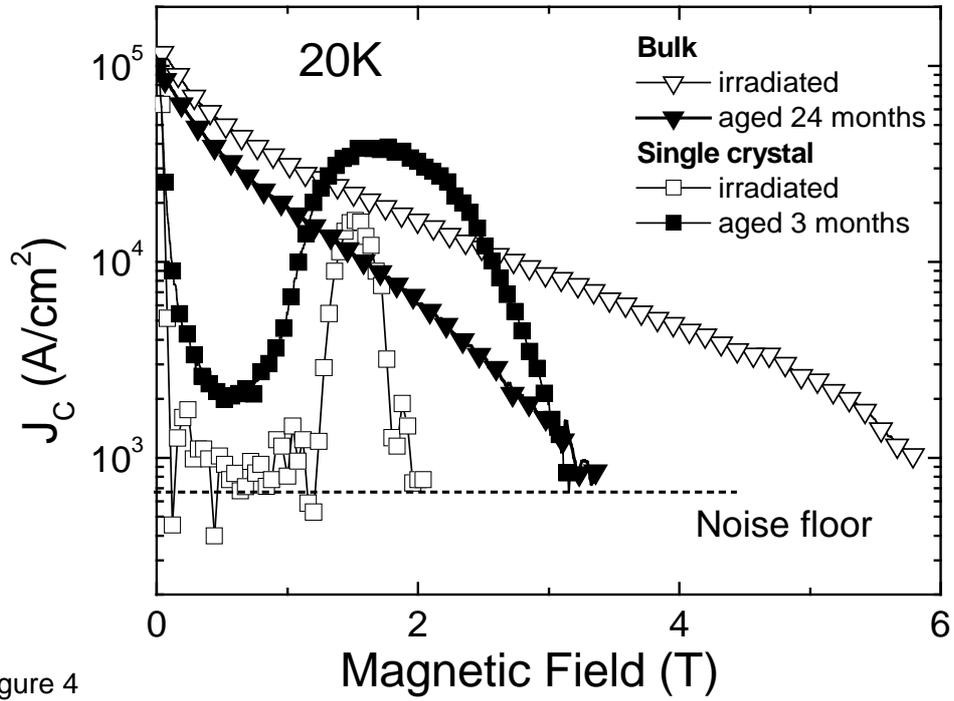

Figure 4